# The Robo-AO-2 facility for rapid visible/near-infrared AO imaging and the demonstration of hybrid techniques

Christoph Baranec*[a], Mark Chun[a], Donald Hall[a], Michael Connelley[a], Klaus Hodapp[a], Daniel Huber[b], Michael Liu[b], Eugene Magnier[b], Karen Meech[b], Marianne Takamiya[c], Richard Griffiths[c], Reed Riddle[d], Richard Dekany[d], Mansi Kasliwal[d], Ryan Lau[d], Nicholas M. Law[e], Olivier Guyon[f], Imke de Pater[g], Mike Wong[g], Eran Ofek[h], Heidi Hammel[i], Marc Kuchner[j], Amy Simon[j], Anna Moore[k], Markus Kissler-Patig[l], and Marcos A. van Dam[m]

[a]Institute for Astronomy, University of Hawaiʻi at Mānoa, Hilo, HI 96720 USA; [b]Institute for Astronomy, University of Hawaiʻi at Mānoa, Honolulu, HI 96822 USA; [c]Department of Physics and Astronomy, University of Hawaiʻi at Hilo, Hilo, HI 96720 USA; [d]California Institute of Technology, Pasadena, CA 91125 USA; [e]Department of Physics and Astronomy, University of North Carolina at Chapel Hill, Chapel Hill, NC 27599 USA; [f]Subaru Telescope, National Astronomical Observatory of Japan, Hilo, HI 96720 USA; [g]University of California, Berkeley, CA 94720 USA; [h]Benoziyo Center for Astrophysics, Weizmann Institute of Science, Rehovot, 76100 Israel; [i]Association of Universities for Research in Astronomy, Washington DC 20004 USA; [j]NASA Goddard Space Flight Center, Greenbelt, MD 20771 USA; [k]Research School of Astronomy and Astrophysics, Australian National University, Canberra, ACT 2611 Australia; [l]European Southern Observatory, Garching 85748 Germany; [m]Flat Wavefronts, 21 Lascelles Street, Christchurch 8022 New Zealand

## ABSTRACT

We are building a next-generation laser adaptive optics system, Robo-AO-2, for the UH 2.2-m telescope that will deliver robotic, diffraction-limited observations at visible and near-infrared wavelengths in unprecedented numbers. The superior Maunakea observing site, expanded spectral range and rapid response to high-priority events represent a significant advance over the prototype. Robo-AO-2 will include a new reconfigurable natural guide star sensor for exquisite wavefront correction on bright targets and the demonstration of potentially transformative hybrid AO techniques that promise to extend the faintness limit on current and future exoplanet adaptive optics systems.

## 1. INTRODUCTION

We are building a robotic laser adaptive optics (AO) system, Robo-AO-2, for the University of Hawaiʻi (UH) 2.2-m telescope. Robo-AO-2 will deliver diffraction-limited images in unprecedented numbers, and be used to rapidly characterize transients, detect changes in monitored objects early in their evolution, and undertake previously infeasible surveys. We will also demonstrate innovative hybrid AO wavefront sensing techniques that extend the brightness limit of high-contrast exoplanet AO systems.

### 1.1 Robo-AO-2

As discussed in the Astro-2010 Decadal survey [1], large area surveys will dominate the next decade of astronomy. Transient surveys such as LSST and space-based exoplanet, supernova, and lensing surveys such as TESS and WFIRST will join the Gaia all-sky astrometric survey in producing a flood of potential discoveries. The NRC's "Optimizing the U.S. Ground-Based Optical and Infrared Astronomy System" [2] stresses the need to characterize these discoveries through several recommendations: Rec#4c) rapid observations of faint transients; Rec#6) continued investment in critical instrument technologies such as adaptive optics; Rec#7) use existing instrument and research programs to support training to build instruments. Additionally, the 2008 U. S. Adaptive Optics Roadmap [3], recognizing the limitations of adaptive optics on large apertures, noted among its recommendations that, "Mid-sized telescopes provide compelling opportunities for world-class science in specialized fields not typically accessible on larger telescopes due to limitations imposed by schedule / observing model and in some cases specialized capabilities."

---

*baranec@hawaii.edu; 1-808-932-2318; http://high-res.org; http://robo-ao.org

Robo-AO-2 will uniquely address these recommendations by combining near-HST resolution across visible and near-infrared (NIR) wavelengths ($\lambda = 400 – 1800$ nm), unmatched observing efficiency, and extensive, dedicated time on the UH 2.2-m. We will enable high-acuity, high-sensitivity follow-up observations of several tens of thousands of objects per year. Robo-AO-2 will also respond to target-of-opportunity events within minutes, minimizing the time between discovery and characterization, and will interleave different programs with its intelligent queue. Robo-AO-2 will be permanently mounted on the UH 2.2-m, will be available year round, will add NIR imaging, tip-tilt correction capability to $m_V$~17 and will enable excellent, <110 nm RMS, image quality on bright, $m_V$<9, objects using a stellar wavefront sensor.

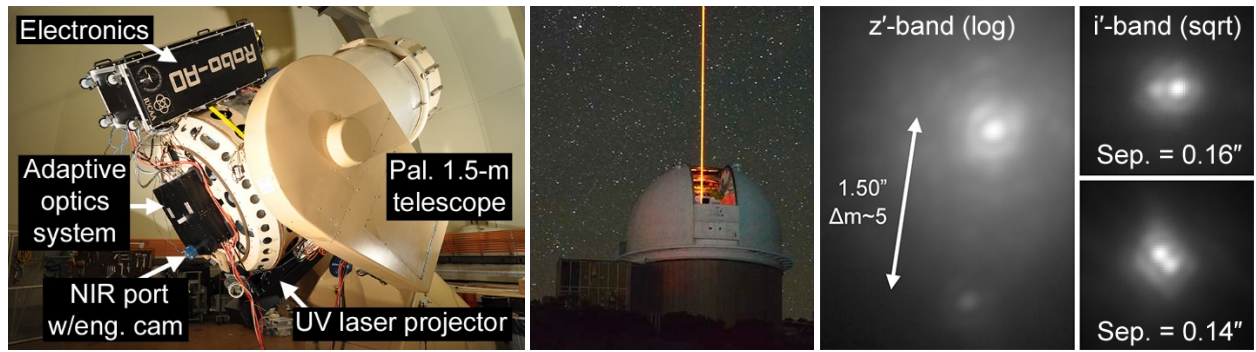

**Figure 1. Left:** The prototype Robo-AO system on the 1.5-m telescope at Palomar. **Center:** The Robo-AO UV laser propagating on sky at the KPNO 2.1-m telescope. **Right:** Robo-AO corrected visible light images of binary stars.

We will base the design of the Robo-AO-2 system on the successful prototype, Fig. 1, which has so far been used for 40 refereed publications [4-43]. The prototype achieved first light at the Palomar 1.5-m telescope in August 2011, first fully automated observing in June 2012 and was decommissioned in June 2015. We moved the system to the Kitt Peak National Observatory 2.1-m telescope where we commenced science operations in November 2015 [44], which lasted until the end of May 2018.

**1.2 Demonstration of hybrid AO technology and techniques**

Scientists are now studying exoplanets using direct imaging and spectroscopy, enabled by second generation AO systems optimized for high-contrast observations around bright stars. Current systems such as the Gemini Planet Imager, Subaru's SCExAO, VLT SPHERE, Palomar P3K, MagAO and LBTAO, have been crucial in the detection and characterization of newly discovered planetary and disk systems. Despite these successes, the target brightness requirement for all these systems ($m_V$<11 to achieve their exquisite correction) precludes routine study of exoplanets around low-mass M dwarfs or nearby young stars. To overcome this limitation, we proposed a hybrid AO solution [45, 46]. By using a brilliant Rayleigh laser to measure the high spatial and temporal order turbulence near the telescope aperture where it dominates, one can extend the faintness limit of natural guide stars (NGS) needed by extreme AO systems. Figure 2 shows the results of hybrid AO simulations for P3K at Palomar and the number of newly accessible targets to exoplanet direct imaging. While low-noise NIR wavefront sensors also have the potential to access this parameter space, they are yet unproven, and will be more effective in combination with hybrid AO techniques once demonstrated. Hybrid AO was first theorized by Ellerbroek in 1994 [47] as one of many ways to extend the useful field of view of laser AO systems. And while hybrid AO enjoyed a successful single test at Starfire Optical Range in the 1990s [48], the US Air Force instead developed multiple Sodium laser AO systems because of their field of view requirements [49].

Robo-AO-2 can implement both laser and natural guide star wavefront sensing in a cost-effective way, making it an ideal platform with which to demonstrate such hybrid AO techniques. It only requires the integration of these data streams in a modern real-time wavefront reconstruction computer to enable a versatile platform for new tomographic investigations. Simulations of the imaging improvement with hybrid AO on the UH 2.2-m, as shown in §5, indicate the visible light Strehl should roughly double at $m_V$~12, and extend equivalent NGS performance to stars ~2 magnitudes fainter.

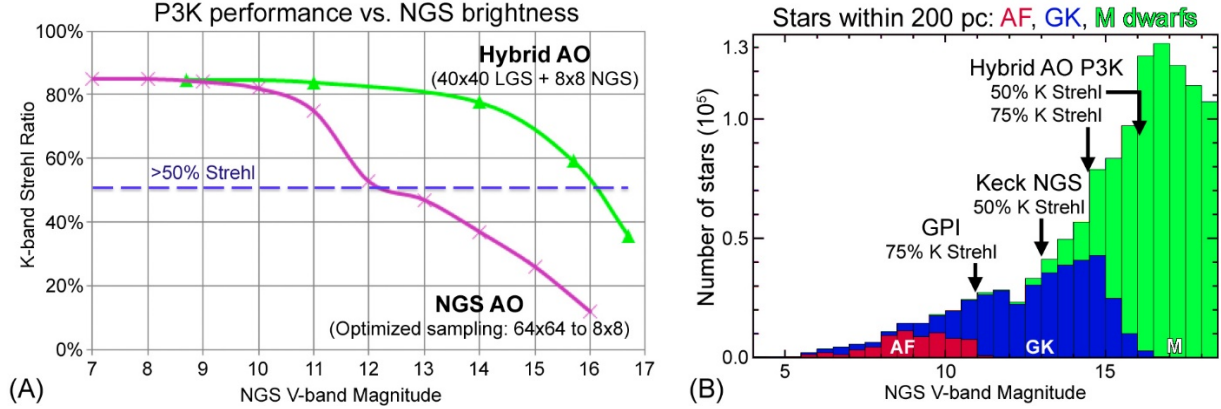

**Figure 2. A:** Our simulations show that hybrid AO on P3K would enable high-contrast imaging on stars ~4 magnitudes fainter than the current NGS only approach. **B:** The new parameter space for direct exoplanet study opened by hybrid AO includes M dwarfs and the youngest early-type stars in associations out to 200 pc. Star counts courtesy B. Bowler using TRILEGAL.

## 2. UNIQUELY ENABLED SCIENCE USING ROBO-AO-2

Robo-AO-2 will enable a broad spectrum of unique and previously infeasible science with its combination of acuity, efficiency and flexible scheduling (see Table 1).

| § | Science Topic | Wavebands | Number and/or Cadence |
|---|---|---|---|
| 2.1 | Wide exoplanets and brown dwarfs | H | 5,000-10,000 targets |
| 2.2 | Transit exoplanet hosts (e.g., TESS) | Broadband vis. | >10,000 targets |
| 2.3 | Asteroseismology and multiplicity | Broadband vis. | Several thousands |
| 2.4 | Discovering/monitoring L. Quasars | i' and H | >25,000 + monitoring 3 nights/mo. |
| 2.5 | Ice Giant monitoring | g' - z' | Snapshot 2-3 times/night |
| 2.6 | IR transient characterization: e.g., novae, WRs, BDs, EM-GW | g' - H | Rapid response, declining follow-up cadence |
| 2.7 | Small body nucleus characterization – Manxes | g' - H | Few night response, ~10 Manx comets per year |
| - | Multiplicity in stellar clusters | z' – H | Several hundred / cluster |
| - | Monitoring Jets/Outflows/Shocks | Narrowband | Several times per year |
| - | Vetting archival surveys for blends | r', i', or J | 300 per year |
| - | Astrometric microlensing | Y, J, or H | Dozens of high cadence events/year |
| - | Stellar Multiplicity from large surveys [42, 50] | Discovery: r', i' | >1,000 (PS1) |
| | | Colors: g' - H | >10,000 (Gaia [51]) |

**Table 1.** A sampling of the breadth of science to be enabled by the new Robo-AO-2 system.

### 2.1 Wide exoplanets and brown dwarfs around young stars

Direct imaging is unique for understanding gas-giant extrasolar planets and is complementary to other methods such as radial velocities, transits, and microlensing. Direct imaging provides photometry and spectroscopy of individual planets' thermal emission that can be used to determine luminosities, temperatures, surface gravity, and photospheric composition (e.g. [52-54]). Moreover, the observed separation and mass distributions can provide key tests for theories of giant planet formation. One of the main surprises from direct imaging studies is the existence of a population of extremely wide (>100 AU) planets around young (1-100 Myr) stars, at separations far beyond the standard operating regime of the core accretion and disk instability mechanisms (e.g. [55-57]). While such objects perhaps do not conform to the standard expectations/biases of a planet, they are a compelling population for detailed study.

We will perform a groundbreaking census of these wide-separation young exoplanets with Robo-AO-2. Previous discoveries of wide exoplanets have occurred in a heterogeneous fashion, with a notional frequency of ~1% [58, 59]. The rarity of these objects requires unparalleled efficiency in order to survey thousands of young stars which is beyond anything previously done or currently planned with conventional high angular resolution platforms (HST or AO). The required flux ratios of ~100-1000, angular separations of ~0.5-10″, and access to the southern hemisphere are all met by Robo-AO-2.

We calculate that a Robo-AO-2 survey will produce a sample of dozens of young gas-giant planets, along with many higher mass brown dwarfs. Such a yield will enable proper statistical measurements of the frequency, mass and separations of this wide population as a function of stellar host mass (e.g. [60, 61]), thereby constraining theories as to their origin. Similarly, with a large sample the connection between wide massive planets (~5-13 $M_{jup}$) and brown dwarfs (>13 $M_{jup}$) can be explored to see if the two mass regimes appear as separate populations or are continuous across the deuterium-burning limit. In addition, all of the discoveries will be readily accessible to photometric and spectroscopic follow-up with conventional AO-equipped telescopes, HST, and JWST. This survey of wide-separation exoplanets around young stars will be complementary to the extreme AO systems, P3K and GPI, as the youngest and lowest-mass stars are too faint for high-order wavefront sensing. Also, wide exoplanets can be followed up with higher S/N and spectral resolution than small-separation discoveries from extreme AO systems. These surveys can be combined to create a more extensive census of the gas-giant planet population from ~3-3000 AU.

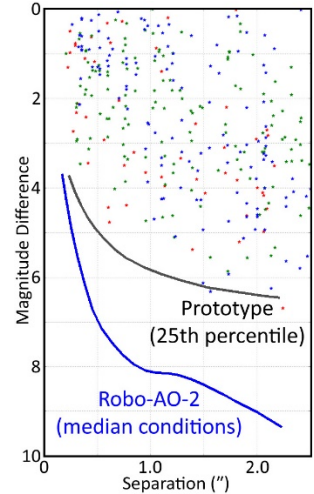

**Figure 3.** Robo-AO-2 will be able to detect ~2 mag. fainter companions than the prototype in a wider range of observing conditions. Individual points represent companions detected as part of our survey of 3,380 Kepler candidate exoplanet host stars [10, 18, 23, 39].

### 2.2 Securing exoplanets and their sizes from transiting planet surveys

NASA's Transiting Exoplanet Survey Satellite (TESS) will detect >5,000 new candidate exoplanets extending down to Earth mass and surrounding a wide variety of host stars (all typically $m_V$<17). Robo-AO-2 will provide the necessary high-resolution imaging to turn these candidates into confident detections and to characterize their stellar host systems. Our Robo-AO-2 follow-up survey will serve three functions that help maximize the return of TESS: 1) vet for false planet detections such as a background eclipsing binary or bound secondary star orbited by a larger planet; 2) search for secondary sources of light in the target aperture that will dilute the transit signal, reducing the measured planet radius; and 3) provide statistics on the rates of planet occurrence as a function of host star multiplicity.

Our Robo-AO-2 survey of TESS and other survey-discovered planet candidates will dramatically improve on previous results from the prototype survey of Kepler objects [10, 18, 23, 39]. For example, Robo-AO-2 has better contrast and an improved inner working angle (Fig. 3) and TESS surveys a closer sample of stars. As a result, Robo-AO-2 will identify secondary light sources at much closer physical distances, e.g., ~2 AU for M dwarfs and ~25 AU for G stars; and is more sensitive to binaries which are most common at separations of ~50 AU [62]. The entire sample of TESS planet candidates accessible from 20° N can be screened for secondary light sources in ~25 nights. The resulting contrast curves will be crucial ingredients for computing false planet probabilities (FPPs; [63]) that are needed to correct planet occurrence estimates. In addition, the unknown multiplicity of the Kepler field (and the TESS and K2 fields) is a significant challenge for planet occurrence calculations. For example, hot Jupiters have a measured occurrence rate of 1% around nearby single stars and 0.5% in the Kepler Field [64]. One theory to explain the factor of two is that hot Jupiters form infrequently in binary systems (except for very wide ones) and half of the Kepler targets are binaries. We will compare the imaged companion rates of more distant Kepler targets with a control sample of nearby field stars and determine if this factor of two correction applies to the 22% measured occurrence of Earth-size planets in the Habitable Zone [65].

### 2.3 Asteroseismology of cool stars

Asteroseismology of solar-type and red-giant stars is one of the fastest growing fields in stellar astrophysics, with nearly 20,000 detections by Kepler alone [66] and an order of magnitude more expected by TESS. While these detections have enabled several breakthrough discoveries [67-69], the limited angular resolution of space photometry missions left many questions unanswered. For example, oscillation amplitudes vary by a factor of 2 among stars with similar fundamental properties, and in some cases are partially or fully suppressed [70]. Understanding these variations is crucial to constrain the poorly understood physics of convection, which stochastically drives and damps oscillations.

We will use Robo-AO-2 to undertake the first comprehensive AO survey of asteroseismic stars, probing the role of undetected companions on the dilution of oscillation amplitudes. The survey will constrain the occurrence rate of wide binary companions (~> 20 AU and ~>100 AU for typical Kepler dwarfs and giants, respectively) as a function of stellar properties of the primary, which are precisely determined through asteroseismology. We will directly measure how the multiplicity rate differs as a function of stellar mass and evolutionary stage. Asteroseismic measurements of spin-axis inclinations combined with multi-epoch imaging for nearby asteroseismic targets will furthermore allow insights into the spin-orbit alignment of binary systems.

### 2.4 Large survey to discover and monitor lensed quasars

We will use Robo-AO-2 to perform the largest-ever search for new gravitational lenses on angular scales below 1.5″. This survey will increase the number of known gravitationally lensed quasars by almost an order of magnitude. A large homogenous sample of such systems will be a unique tool to study key issues including cosmology and sizes of quasar accretion disks. Strong lensing can also be used to constrain galaxy properties including galaxy mass distribution, mass evolution, dark matter content, and mass-to-light ratios and evolution, as well as the size and structure of quasar accretion disks: this latter measurement can be uniquely addressed using microlensing, whereby a lensed qusaar image is additionally microlensed by stars in the lensing galaxy. When followed up using X-ray observations, microlensing studies lead to measurement of the size of the innermost accretion disk and the x-ray reflection component dominated by general relativistic effects, and the spin of the black hole.

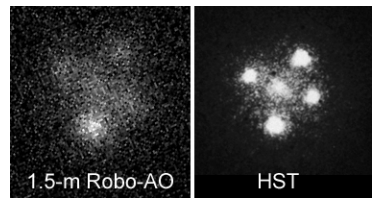

**Figure 4.** The lensed quasar G2237+0305, 1.6″ across. While challenging for the 1.5-m Robo-AO system, the increased aperture and Strehl ratio (each roughly ×2) at the 2.2-m telescope will make discovering and monitoring such sources routine.

Ninty percent of all systems will have angular separations smaller than about 1.5″, which require high angular resolution. Robo-AO-2 will perform diffraction-limited H-band imaging of all quasars with $z > 1$ and $i < 20.0$ mag for which a suitable tip-tilt star is present (approximately 16% at a 30º galactic latitude), and 2-3× sharper-than-seeing imaging of quasars without a tip-tilt star. H-band imaging is optimal for detection of the lensing galaxy. To have an excellent chance of detecting most components of likely lens geometries (Fig. 4), Robo-AO-2 will require approximately 5-min integrations. Selecting likely targets from the seeing-limited SDSS survey will yield around 25,000 candidate targets. We will augment this target list with objects from the PanSTARRS 3pi survey, which no other group has observed, and a brighter magnitude cut can be employed, increasing chances of success. Observing ~90 candidates per night, this survey will be completed in ~9 months spread over several years. We estimate that this search will yield 300-700 new lensed quasars suitable for time delay measurements.

Lens monitoring in the i′-band will enable measurement of time delays of large sample of known lenses. The follow-up monitoring program will require approximately an additional 3 nights per month, selecting the brightest of the newly discovered lenses for increased signal-to-noise (S/N).

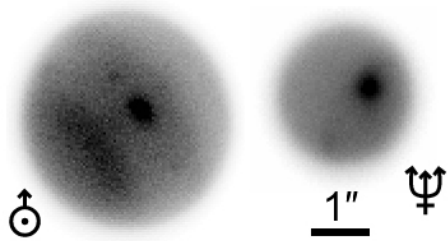

**Figure 5.** Prototype Robo-AO images showing storm activity on Uranus and Neptune. Robo-AO-2 will provide higher Strehl and 50% sharper routine imaging of the ice giants.

### 2.5 Monitoring of Ice Giants

Although Keck and HST have obtained the highest-spatial-resolution imaging of Uranus (3.8″) and Neptune (2.3″) from Earth, time on these facilities is extremely limited. That significantly limits our ability to understand the full life cycle of major storms on these dynamic planets including burning questions from the Voyager mission: Despite their many similarities, why is it that Neptune has an internal heat source but Uranus does not? What are the origins, evolution, and destruction of dark vortices? These puzzles are likely related to the style of convection, circulation and dynamics in giant planet atmospheres, where large superstorms can provide a significant fraction of the planet's radiant emission [71]. Models suggest that convection becomes more intermittent with increasing heavy-element abundance [72]. This trend that can be tested by monitoring the storm activity regularly with Robo-AO-2, constraining the nature of convection and tracing the origins of storm systems on these planets.

Both the Planetary Science Decadal Survey [72] and the Panel on Solar Wind-Magnetosphere Interactions Panel from the Heliophysics Decadal Survey [74] have identified a Neptune/Uranus Orbiter and Probe as the next priority flagship mission. Mission planning activities will be informed by our current understanding of the planet. Regular wide-wavelength monitoring with Robo-AO-2 will help inform instrument choices and guide science priorities.

## 2.6 Rapid high-S/N follow-up of IR Transient Surveys

Ground-based, optical time-domain transient surveys are blind to transients and eruptive variables that are self-obscured or deeply embedded in dusty environments like high mass star-forming regions. The infrared provides a crucial probe for transients enshrouded by dust with high optical extinction. Pursuing systematic observations of the largely un-explored dynamic infrared sky is therefore essential for identifying and characterizing the signatures of red transients and variables.

For example, Gattini-IR [75] is a J-band optimized telescope at Palomar that will survey the entire accessible sky every night to a depth of 16.4 AB mag with a ~9″ plate scale. Robo-AO-2 will be the ideal Gattini-IR follow-up machine given its ability to perform rapid high S/N and diffraction-limited follow-up observations of interesting IR transients and variables. The Gattini-IR team will use Robo-AO-2 to obtain additional observations of (from lowest to highest frequency): classical novae, heavily obscured supernovae, dusty Wolf-Rayet stars, weather related activity from brown dwarfs, stellar merger candidates, and possible electromagnetic counterparts of neutron star mergers [76, 77]. Here, we highlight the cases of classical novae and electromagnetic/gravitational wave events.

The field of nova research is in the middle of a revolution. The discovery by the Fermi satellite that many classical novae produce $10^{35}$ - $10^{36}$ erg/s of gamma-ray emission [78] has revealed that shocks are important in novae. Shocks may even explain the long-standing conundrum of how dust can form in the hostile environment of a nova. The "shock-dust" model [79] predicts that dust will form in a toroidal morphology behind a slow-moving radiative shock that was initially responsible for the gamma-ray emission. In novae, dust production is determined from an IR excess, and it may or may not be accompanied by deep V-band dust dips. The occurrence of V-band dips will depend on the nature of the dust formation and the viewing angle of the system. Gattini-IR will perform an IR census of dust producing novae with an estimated rate of 20-50 novae per year [80], whereas the currently observed optically-selected rate is ~8 per year. The robotic nature of Robo-AO-2 is ideal for rapid follow-up of novae detected by Gattini-IR. Simultaneous optical/IR (g' i', J, H) imaging of novae outbursts with ~2-day cadences over 3 months will allow us to identify dust production and V band dips. The spatial resolution of Robo-AO-2 will be crucial to distinguish novae in Gattini-IR's 9″ pixels. The Gattini-IR/Robo-AO-2 nova sample will span a wide-enough range of ejecta velocities (as indicated by optical spectra and/or the speed class) and inclinations to achieve our goal of verifying the "shock-dust" model.

Currently, a major obstacle of performing electromagnetic follow-up on gravitational wave events detected by Advanced LIGO is the coarse localization of hundreds of square degrees [81]. Gattini-IR will pinpoint the IR counterpart of a neutron star merger if the source is brighter than 20 AB (J band; 5-σ), assuming a median localization of 250 deg$^2$ and a 5-night exposure. Robo-AO-2 increases the point-source sensitivity of the 2.2-m in the infrared, making EM-GW event light curves detectable down to $m_J$=23 AB with a S/N of 10 in ~30 minutes. The rapid follow-up capabilities of Robo-AO-2 are essential due to the fast decay rates predicted by theoretical models [82]. We require full optical/NIR (g', r', i', J, H) imaging with Robo-AO-2 at rapid cadences (several times a night) for the first 5 days after the transient and lower ~2 day cadences for the following month in order to verify the SED and light curve as a Neutron Star merger. The 3-hour difference between Palomar and Maunakea is ideal for performing swift follow-up with Robo-AO-2 of Gattini-IR transients discovered in the same night. Results from this study will allow us to address the question of whether neutron star mergers are the production sites of *r*-process elements such as gold, platinum and other heavy elements in the Universe.

## 2.7 Small body nucleus characterization – Manxes

The PanSTARRS and Catalina all-sky surveys have been discovering new types of small bodies that can shed light on the process of the formation of habitable planets. Object on nearly parabolic long-period (LP) comet orbits are usually rich in volatiles and are strongly outgassing at their time of discovery as they approach perihelion. "Manx" comets are objects on LP comet orbits that are nearly inactive, even at perihelion [83]. They may represent relatively dry inner solar system material that was tossed to the outer solar system during the era of planet formation, and as such may be useful to distinguish between dynamical models of solar system formation. We need to characterize the surface mineralogy with g', r', i', z', Y, J, and H photometry. Unfortunately, these are discovered typically when they are brightest, and they fade very rapidly as they move away from perihelion. Often they are lost because they become too faint by the time they make it to the top of the priority for queue scheduled telescopes. If we could observe these *immediately* after discovery when they are still bright we could more than double the rate of characterization. Robo-AO-2 will allow for rapid turn-around observations and the ability to go fainter on the UH 2.2-m. Sensitivity is significantly increased in the background-limited near-infrared (Y, J, H; [84]) and visible photometry can be acquired simultaneously with the near-infrared observations.

This capability will be extremely valuable for characterizing the next interstellar objects (ISO) to be discovered by PanSTARRS, as 1I/2017 U1, the first ISO dropped by a factor of 100 in brightness in less than a month after discovery significantly limiting the science that could be done [85].

## 3. DESCRIPTION OF THE INSTRUMENT

The design of the Robo-AO-2 system is based on the successful prototype [4, 5] and makes several improvements to the performance and capability. Robo-AO-2 will comprise a UV laser projector, a Cassegrain mounted adaptive optics system (Fig. 6) with low-noise, high-speed visible and infrared imaging arrays that double as tip-tilt sensors (see Table 2), a new reconfigurable stellar wavefront sensor, and a set of electronics and control computer with additional functionality.

| Detector | Format | Field | Pixel Scale | Read Noise | Full frame rate | Tip-tilt rate | Initial filters |
|---|---|---|---|---|---|---|---|
| EMCCD | $1024^2$ | 31″×31″ | 0.030″ | <1e- | up to 26 Hz | to 500 Hz | g', r', i', z', Hα, SII, OIII, LP600 |
| SAPHIRA | 320×256 | 17″×14″ | 0.055″ | <1e- | up to 400 Hz | to 8 kHz | Y, J, H, FeII, Y+J+H |

**Table 2.** The Robo-AO-2 imaging/tip-tilt detectors will enable high acuity imaging over λ=400-1780nm.

### 3.1 Cassegrain mounted AO system

The adaptive optics system and science cameras will reside within a Cassegrain mounted structure of approximate dimensions 1 m × 1 m × 0.2 m (Fig. 6). Light from the telescope will enter the instrument and be intercepted by a fold mirror which directs a 2 arc minute diameter field to a dual off-axis parabolic (OAP) mirror relay. The first fold mirror will be on a linear motorized stage that can be moved out of the way to enable a 12 arc minute, 75 mm, pass-through mode for seeing-limited instruments to be co-mounted with AO. This will eliminate the need to ever remove the AO system and is made possible by the 2.2-m telescope nominal 313 mm back focal distance. Also with the fold mirror moved, a calibration source that matches the 2.2-m telescope focal ratio and exit pupil position will simultaneously mimic the ultraviolet laser focus at 10 km and a thermal source at infinity.

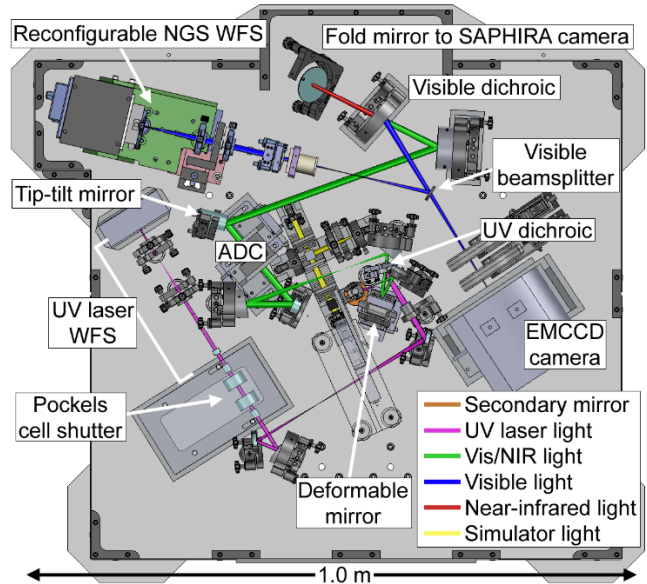

**Figure 6.** The layout of the Robo-AO-2 system will be similar to that of the prototype with room for the new reconfigurable NGS WFS and visible beamsplitter mirror.

The first OAP will image the telescope pupil onto the deformable mirror. After reflection off the deformable mirror, the UV laser light will be selected off with a dichroic mirror and sent to a UV optimized wavefront sensor. The visible and infrared light will pass through the dichroic and will be refocused by another OAP. The light will then be relayed by a second OAP relay which includes a tip-tilt corrector and an atmospheric dispersion corrector (ADC; 400nm < λ < 1.8μm). The final relay element creates a telecentric F/41 beam then split by a dichroic mirror at λ = 950 nm.

### 3.2 Science cameras / tip-tilt sensors

We will use an Andor iXon Ultra 888 camera for our visible channel. This camera uses the same E2V CCD201 electron-multiplying CCD as the prototype Robo-AO Andor 888 camera but with two major improvements; new electronics are optimized for fast readout, so the cameras can read out the full frame at 26 Hz as opposed to 8.6 Hz – providing better post-facto image motion correction on brighter targets. A new USB 3.0 interface obviates the need of custom interface cards in the control computer.

We will use a Selex Avalanche PHotodiode InfraRed Array (SAPHIRA) detector for the near infrared channel. These were originally developed by Selex ES and ESO for the GRAVITY instrument on the VLTI [86] and have been adopted by ESO for future AO wavefront sensing applications. We at the Institute for Astronomy have been funded by the NSF to further develop these 320×256 devices for infrared tip-tilt and wavefront sensing and by NASA for space applications (and also now larger formats) as they are capable of very fast readouts, >400 Hz full frame with 32 parallel outputs, and very low noise, <1e-, via avalanche amplification of the signal in the HgCdTe before being read out. In summer 2014, we tested an early device as a simultaneous tip-tip sensor and science camera where we captured simultaneous data with both the SPAHIRA and EMCCD while tip-tilt guiding with the SPAHIRA at 100 Hz [6]. One of the more recent science grade SAPHIRA detectors (AR-coated, sensitive from 1-1.8 μm, and with only a few bad pixels) was being used with Robo-AO at Kitt Peak [87]. The SAPHIRA cryostat is able to hold larger format detectors so we can install new detectors when they become available.

### 3.3 Larger format, lower noise, faster readout UV laser wavefront sensor

The prototype Robo-AO system uses an 11×11 rectilinear Shack-Hartmann wavefront sensor (WFS). The detector is an E2V UV optimized CCD39 ($80^2$ pixels; 72% quantum efficiency at 355 nm). We binned the pixels by a factor of 3 (binned read noise of $6e^-$ at 1.2 kHz) and used 2×2 binned pixels to calculate the slope of each subaperture. For Robo-AO-2, we will maintain the projected subaperture size on the telescope primary mirror, increasing the pupil sampling to 16×16 subapertures. We will use an ultraviolet optimized Nüvü Hnü128$^{AO}$ EMCCD detector ($128^2$ pixels; <0.1e- read noise at 20MHz pixel rate). The detector is an E2V CCD60 with an Astro Broadband AR coating that has >50% QE at the laser wavelength. We will bin the pixels by a factor of 2 and use 4×4 binned pixels to calculate the slopes in each subaperture, leading to a more linear response than previously achieved with quad cells. We will operate at a faster 1.8 kHz binned full-frame rate. Even with the greater number of pixels being read, and the faster frame rate, the EM gain allows us to maintain a higher S/N slope measurement with Robo-AO-2.

Range gating of the pulsed laser, to block Rayleigh scattered light outside the focused beam waist, will be accomplished as implemented in the prototype system - with a Beta Barium Borate Pockels cell optical switch.

### 3.4 New stellar wavefront sensor for bright targets and hybrid AO demonstration

To support the hybrid AO demonstrations we will add a reconfigurable NGS wavefront sensor to the AO system. We will copy the design for the P3K wavefront sensor [88]. The P3K wavefront sensor uses a reflective collimator to image the pupil, a lenslet array exchanger to select between 64-, 32-, 16- and 8-across pupil sampling modes (as well as an open pupil imaging mode) and an optical relay and detector on a focus stage to adjust to the different lenslet array focal lengths. The lenslet arrays are mounted in individual lens tubes that can be swapped without disturbing the optical alignment of the wavefront sensor.

The Robo-AO-2 NGS wavefront sensor will use a standard Nüvü Hnü128$^{AO}$ EMCCD detector with a Basic Midband AR coating. The detector achieves an unbinned full-frame rate of 1.0 kHz. We will populate the lenslet array exchanger with lenslet arrays that sample 16, 8, 4 and 2 across the pupil, along with a central lens to enable additional tip-tilt guiding. The 16 across mode will be used during robotic science operations when guiding on bright targets, $m_V<9$, and will have a subaperture size of 4.2″ (over 4 pixels) to enable NGS guiding on Uranus and Neptune. The 8, 4 and 2 across lenslet arrays will be used with the hybrid AO demonstrations. The NGS WFS will be fed by a remotely actuated and selectable 90/10 or 50/50 beamsplitter in front of the visible science camera.

Large telescope laser AO systems require the use of a low-bandwidth NGS wavefront sensor to calibrate non-common path errors between the science instrument and laser wavefront sensor for each new observation, e.g., approximately 30-60s with Keck laser AO. As evidenced by the effective PSF subtraction routines we use on full nights' worth of data, these non-common path errors change slowly over the course of several nights with Robo-AO, likely due to the compact and stiff structure. With the new NGS wavefront sensor, we will implement a single non-common path calibration at the beginning of each night to minimize the effect of non-common path error in the laser wavefront sensor.

### 3.5 On-axis, focus-stable, UV laser guide star

We will copy the prototype laser projector which has a compact, 1.5 m × 0.4 m × 0.25 m, ~70 kg, enclosure. The projector will attach to the side of the UH 2.2-m telescope. Inside the projector will be a higher power (15W) pulsed laser (λ=355nm; Lumentum Q301-HD-1000R HE); a redundant safety shutter; and an uplink tip-tilt mirror to both stabilize the apparent laser beam position on sky and to correct for up to 2′ of differential telescope flexure. A bi-convex lens on an adjustable focus stage expands the laser beam to fill a 15 cm output aperture lens, which is optically conjugate to the tip-tilt mirror. The laser light will be focused to a 10 km line-of-sight distance with a longer 667-m range-gate to compensate for the higher elevation.

The UV laser has been approved for use without human spotters by the Federal Aviation Authority because it is invisible to humans and does not present a hazard for <10 s exposures. Coordination with U.S. Strategic Command is necessary to avoid illuminating critical space assets. Instead of requesting clearance for individual targets (as all other US laser systems do), we adopt the procedure developed for the prototype system and will request open times for azimuth and elevation ranges of 2.5°×2.5° over the entire sky above 30° elevation. This ensures that at any given time there are targets available to observe with the laser and also allows us to observe new targets on the fly, e.g., transients, targets of opportunity, supernovae, etc.

Two improvements will be implemented for the UH 2.2-m laser system. A periscope will be added to the end of the telescope tube to jog the beam on-axis to reduce perspective elongation of the laser - necessary with the extended range-gate. While the laser projector pointing is stable to the level correctable by the uplink tip-tilt mirror, the periscope will include active pointing to compensate for deterministic mechanical flexure. We will also implement passive and active controls on the laser projector focus to compensate for the temperature dependent focus drift found with the prototype. E.g., by using a carbon fiber, as opposed to aluminum, breadboard; and setting the focus position by temperature.

### 3.6 Higher actuator count deformable mirror

To match the laser wavefront sensor, we will upgrade the deformable mirror from a Boston Micromachines 12×12 actuator continuous face-sheet mirror to a 492 actuator version (24 actuators across a circular aperture), with the same 3.5 μm stroke. The new deformable mirror has a total latency and response of <50 μs, approximately half of the previous mirror. We will similarly adopt a Fried geometry for wavefront correction by using a subset of actuators spanning 17 actuators across the pupil.

### 3.7 Robotic software

The Robo-AO-2 system will reuse the robotic software developed for the prototype (fully detailed in [89]). A single computer commands the AO system, the laser guide star, visible and near-infrared science cameras, the telescope, and other instrument functions. Recent optimizations of the code and new power switching hardware have reduced the setup time for each observation to a mere 30-40 seconds (excluding the telescope slew).

The intelligent queue is able to pick from targets in a directory structure organized by scientific program, with observation parameters defined as .xml files. All targets from all programs are initially considered "available" and the queue uses an optimization routine based on priority, slew time and cutoffs to determine the next target to observe.

To have the Robo-AO-2 system work as the prototype does now, we will write interfaces for the new pieces of hardware: the visible, and wavefront sensor cameras; the deformable mirror; and the UH 2.2-m telescope control system. An existing fast-frame rate visible light camera data reduction daemon currently reduces all data shortly after it is acquired. We will create new automatic data reduction daemons to reduce data from the infrared camera, as well as handle long-integration data from each of the cameras. We will be developing additional extensions for the intelligent queue so that it can poll the Maunakea weather station, seeing monitor and laser traffic control system [90] as part of its decision making process. We will work with the Maunakea Laser Operators Group (which includes members from the visible and infrared observatories on Maunakea) on the safe use of the UV laser. We will also be developing a protocol for targets-of-opportunity to be remotely and automatically added into the queue at high priority, e.g. transients discovered by Gattini-IR and other time domain astronomy surveys.

## 4. ADAPTIVE OPTICS PERFORMANCE

We have maintained detailed error budgets for the expected adaptive optics performance of the prototype Robo-AO system under different observing conditions and have validated the performance on sky [91]. This error budget, originally developed by R. Dekany, et al., was validated against on-sky performance of the Keck and 5-m Hale laser AO systems. Using this same tool, we have estimated the performance of the Robo-AO-2 system assuming an $m_V=17$ M type tip-tilt star sensed by either the EMCCD or SAPHIRA cameras with a broad filter (Table 3). The error budgets use measured Maunakea $C_n^2(h)$ profiles derived from a combination of the Gemini ground-layer study [92] and an analysis led by M. Chun, of data from the Maunakea summit MASS/DIMM seeing monitor running since fall 2009. An additional 0.44″ of dome seeing has been added to the $C_n^2(h)$ profiles. We assume a telescope throughput of $0.70^2 = 0.49$ at $\lambda = 355$ nm, and a detector QE of 50%.

Sky coverage for imaging objects that are too faint to be used as tip-tilt guide sources will be modest (although the Robo-AO-2 will be able to guide on stars almost a magnitude fainter than the prototype due to the larger aperture). At a galactic latitude of 30°, by guiding on a $m_V$=17 star in the visible, 16% and 9% of the sky can be accessed with the visible and infrared cameras, respectively, with up to an additional 90 and 80 milli-arcsec (mas) of RMS tip-tilt (two-axis) error to be added in quadrature. Guiding on a $m_V$=17 M type star in the infrared will allow access to 9% and 5% of the sky on the visible and infrared cameras, at the cost of up to 80 and 60 mas of RMS tip-tilt error respectively.

For bright targets that can be used as their own guide star, $m_V$<9, wavefront correction will be exquisite. For example, when observing Neptune in unfavorable 75% seeing conditions at a zenith angle of 35°, Robo-AO-2 will be able to correct wavefront aberrations to the level of 109 nm RMS. This corresponds to Strehl ratios of 11%, 25%, 38% and 49% across the visible g'-, r'-, i'-, and z'-bands.

| Robo-AO-2 Percentile Seeing | 25% | 50% | 50% | 75% |
|---|---|---|---|---|
| Atmospheric $r_0$ at Zenith | 22.1 cm | 16.8 cm | | 10.3 cm |
| Effective seeing at Zenith angle (inc. dome seeing) | 0.70" | 0.83" | 0.95" | 1.17" |
| Zenith angle | 15 degrees | 15 degrees | 40 degrees | 40 degrees |
| **High-order Errors** | Wavefront Error (nm) | | | |
| Atmospheric Fitting Error | 40 | 45 | 51 | 61 |
| Bandwidth Error | 29 | 33 | 37 | 42 |
| High-order Measurement Error | 37 | 41 | 42 | 50 |
| LGS Focal Anisoplanatism Error | 74 | 102 | 118 | 164 |
| Other High-order Errors | 59 | 60 | 71 | 74 |
| **Total High Order Wavefront Error** | 113 nm | 137 nm | 157 nm | 201 nm |
| **Tip-Tilt Errors** | Angular Error (milli arcseconds - mas) | | | |
| Tilt Measurement Error | 11 | 11 | 13 | 14 |
| Tilt Bandwidth Error | 8 | 9 | 8 | 10 |
| Other Tip-Tilt Errors | 6 | 7 | 11 | 10 |
| **Total Tip/Tilt Error (one-axis)** | 15 mas | 16 mas | 19 mas | 20 mas |
| **Total Wavefront Error (NIR TT)** | 125 nm | 148 nm | 166 nm | 209 nm |
| **Total Wavefront Error (Visible TT)** | 127 nm | 150 nm | 168 nm | 212 nm |

| Spectral Band | λ | λ/D | Strehl | FWHM | Strehl | FWHM | Strehl | FWHM | Strehl | FWHM |
|---|---|---|---|---|---|---|---|---|---|---|
| g' | 0.47 μ | 0.044" | 6% | 0.06" | 2% | 0.07" | 1% | 0.16" | 0% | 0.50" |
| r' | 0.62 μ | 0.058" | 17% | 0.07" | 9% | 0.07" | 5% | 0.08" | 1% | 0.17" |
| i' | 0.75 μ | 0.070" | 29% | 0.08" | 18% | 0.08" | 12% | 0.08" | 4% | 0.10" |
| J | 1.25 μ | 0.117" | 62% | 0.12" | 52% | 0.12" | 45% | 0.13" | 29% | 0.13" |
| H | 1.64 μ | 0.153" | 75% | 0.16" | 68% | 0.16" | 62% | 0.16" | 48% | 0.17" |

**Table 3.** The superior Maunakea atmospheric statistics and higher spatiotemporal order adaptive correction combine to reduce the effective wavefront error in median conditions by nearly 110 nm in quadrature over the prototype. Strehl in the i'-band nearly doubles in median conditions when observing near Zenith. Other high-order and tip-tilt errors include chromatic, scintillation, aliasing, calibration and digitization errors.

## 5. HYBRID AO DEMONSTRATIONS WITH ROBO-AO-2

### 5.1 Simulation of hybrid AO

In previous work [45, 46], we investigated the performance of hybrid AO with P3K on the 5-m Hale telescope. The P3K hybrid AO concept uses a bright Rayleigh laser guide star augmented by a low-order natural guide star. By measuring more than just tip-tilt using the NGS, we can reduce the effect of focal anisoplanatism inherent in LGS wavefront sensing. Focal anisoplanatism (FA) error from the laser arises primarily from the perimeter of the telescope pupil where the geometric error is large; interior to the pupil there exists little geometric discrepancy due to finite beacon altitude. The hybrid AO wavefront sensing approach uses the low-order NGS to fill in the missing information at the edge of the pupil.

In Figure 7, we present a hybrid, NGS and LGS simulation of Robo-AO-2 at the UH 2.2-m in median seeing. The Robo-AO-2 hybrid AO mode takes advantage of the better and lower relative altitude seeing on Maunakea along with the reduced focal anisoplanatism from a smaller telescope. The simulation is performed with '*yao*', a mature AO performance simulation package widely used in the AO community [93-95]. *Yao* uses Monte Carlo physical wave-optics propagation through turbulence to simulate realistic AO performance, capturing fitting, measurement, bandwidth/control and tomographic/FA errors. The simulations assume a slightly different configuration than that as described in §3 and §4 due to a change in available laser WFS cameras: a 19×19 laser WFS, reconfigurable (16×16, 8×8, 4×4 or 2×2) NGS WFS observing an A0 type star, a 20×20 actuator deformable mirror and a tip-tilt mirror. Each wavefront sensor uses a zonal regularized least-squares reconstructor, except in the cases of the 2×2 WFS, where a low-order Zernike modal reconstructor was used. A tomographic reconstructor that takes into account the $C_n^2(h)$ is used for hybrid AO correction. A simple integral controller with a 2-frame delay is used in all simulations. A total of two seconds of integration time, corresponding to 3000 iterations were simulated. The frame rate of the NGS WFS was optimized for each magnitude by varying it from 23 Hz to a maximum of 1500 Hz.

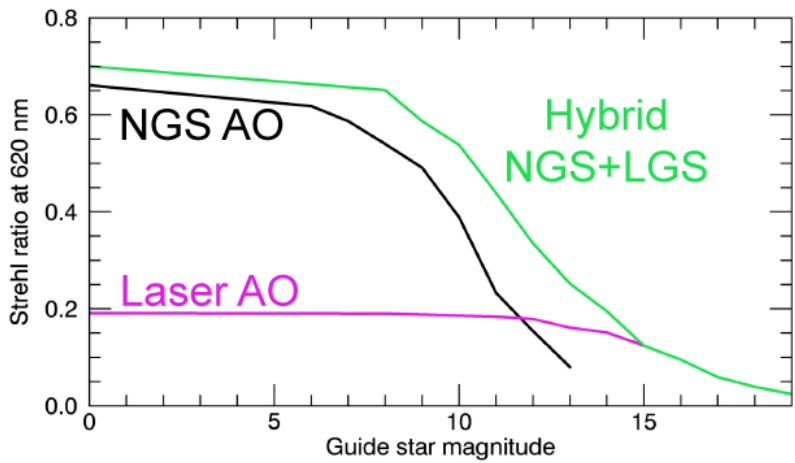

**Figure 7.** The Robo-AO-2 hybrid AO combination of LGS and NGS signals double the achievable λ=620 nm Strehl ratio at V~12, where the NGS and LGS AO performance are nearly equal. Hybrid AO also enables similar levels of correction for NGS approximately 2 magnitudes fainter than otherwise possible.

While not as dramatic as the hybrid AO improvement seen for P3K (Fig. 2), much of this can be attributed to the smaller FA error at the 2.2-m and better seeing conditions at Maunakea. We expect that the simulation results, and subsequently on-sky results, could be improved by exploring other tomographic reconstruction and control loop strategies, but even in this simplified case, hybrid AO approaches will make a significant improvement over traditional NGS or LGS AO strategies.

**5.2 On-sky demonstration**

We will start the demonstrations of hybrid AO in early 2020, shortly after the Robo-AO-2 system is fully commissioned. A strength of evaluating hybrid AO techniques with Robo-AO-2 is that we can manually interrupt the intelligent queue when atmospheric conditions match those which we desire to test. We can also easily change how the AO system operates, including reconstructors and control laws, by merely editing the system's text configuration files. We will be able to update the tomographic reconstructors with the local MASS/DIMM turbulences profiles. Robo-AO-2 already incorporates the use of asynchronous WFS data (as evidenced by our IR tip-tilt demonstration [6]) and the addition of the NGS WFS will not overtax the AO loop that currently uses <3% of a single CPU core during operation.

The efficacy of hybrid AO vs. NGS and LGS AO will be assessed using images captured with the science cameras (including 400Hz speckle evolution statistics with the NIR camera) and with the AO loop telemetry which can record wavefront slopes and deformable mirror actuator commands at each step.

**5.3 Impacts on existing large telescope AO systems**

The demonstrations of hybrid AO with Robo-AO-2 will have a potentially transformative impact on many current and future large telescope AO systems. FA error of ~140 nm RMS on 8 to 10-m Sodium LGS AO systems on Maunakea [96] is slightly worse than for the Robo-AO-2 Rayleigh LGS, suggesting that hybrid AO will be more effective on the larger apertures. Extreme AO systems such as GPI (North), SCExAO and the planned Keck Planet Imager and Characterizer [97] will extend their NGS faintness limit when combined with their second generation Sodium LGS [98]. Traditional Sodium laser AO systems can reduce the effects of FA error by implementing hybrid AO with a higher-order NGS WFS instead of just the required NGS tip-tilt-focus sensors. While multi-NGS tomographic reconstruction has been demonstrated to match simulations (e.g., VLT's MAD [99]), it is an outstanding problem that multi-laser tomography systems do not reach their expected potential performance [100]. By examining the response of tomographic reconstruction in hybrid AO, we simplify the problem (one vs. many LGS) which may shed light on our deficiencies of understanding of multi-laser tomography.

## 6. DISCUSSION

Robo-AO-2 will combine near-HST resolution across visible and near-infrared wavelengths with unmatched observing efficiency, and extensive, dedicated time on the UH 2.2-m. This will enable high-acuity, high-sensitivity follow-up observations of several tens of thousands of objects per year. It will also respond to target-of-opportunity events within minutes, minimizing the time between discovery and characterization, and will interleave different programs with its intelligent queue. With the addition of its reconfigurable stellar wavefront sensor, Robo-AO-2 is also an ideal platform with which to demonstrate such hybrid AO techniques that will be critical to extending the faintness limits of current and future exoplanet adaptive optics systems.

## ACKNOWLEDGEMENTS

C.B. acknowledges support from the Alfred P. Sloan Foundation. The Robo-AO-2 system is supported by the National Science Foundation under Grant No. AST-1712014, the State of Hawaii Capital Improvement Projects, and by a gift from the Lumb Family. Support for the infrared camera for Robo-AO and Robo-AO-2 was provided by the Mt. Cuba Astronomical Foundation and through the National Science Foundation under Grant No. AST-1106391.